\documentstyle[12pt]{article}

\catcode`@=11
\def\secteqno{\@addtoreset{equation}{section}%
\def\theequation{\thesection.\arabic{equation}}}
\catcode`@=12
\secteqno
\newcommand{\be}{\begin{equation}}
\newcommand{\ee}{\end{equation}}
\newcommand{\bea}{\begin{eqnarray}}
\newcommand{\eea}{\end{eqnarray}}
\newcommand{\bref}[1]{(\ref{#1})}
\newcommand{\eps}{\epsilon}
\newcommand{\veps}{\varepsilon}

\newcommand{\D}{\delta}\newcommand{\ep}{\epsilon}\newcommand{\A}{\alpha}
\newcommand{\pa}{\partial}
\catcode`@=11
\newcommand{\vs}{\vskip 6mm}

\begin{document}
\vfill
\vbox{
\hfill February, 2001 \null\par
\hfill 
}\null

\vskip 10mm
\begin{center}
{\Large\bf Gauge Transformations and Weak Lax Equation
\\
}\par
\vskip 10mm
{Takeshi FUKUYAMA$^\dagger$,  
Kiyoshi KAMIMURA$^\ast$ and Kouichi TODA$^\dagger$}\par
\medskip
$^\dagger$Department of Physics, Ritsumeikan University,\\ 
Kusatsu, Shiga, 525-8577 Japan\\
$^\ast$Department of Physics, Toho University, Funabashi, 
274-8514 Japan 
\medskip
\vskip 10mm
\end{center}
\vskip 10mm
\begin{abstract}
We consider several integrable systems from a standpoint of the 
SL(2,R) invariant gauge theory. In the Drinfeld-Sokorov gauge, 
we get a one parameter family of nonlinear equations from zero 
curvature conditions. For each value of the parameter the equation 
is described by weak Lax equations.
It is transformed to a set of coupled equations which pass the 
Painlev\'{e} test and are integrable for any integer values of the 
parameter. 
Performing successive gauge transformations (the Miura transformations) 
on the system of equations we obtain a series of nonlinear equations. 
 
\end{abstract}

\vskip 2cm

\section{Introduction}
\indent

Integrable systems in (1+1) dimensions have been discussed extensively 
so far and various techniques to analyze the integrability
have been developed using the Lax pairs, the inverse scattering method, 
the Hirota's direct method and the Painlev\'{e} test etc.\cite{das}. 
However, these methods are mainly restricted in (1+1) dimensions.
In (1+1) and higher dimensions, there is a well known conjecture by Ward 
\cite{ward} that any integrable systems are derived from (anti) self-dual 
Yang-Mills gauge theories.
So it is very important to re-examine integrable systems from gauge 
theoretical view points.
The term {\it gauge} has been often used 
with several different meanings in different papers and textbooks 
in the world of nonlinear mathematical physics.
When we argue gauge transformations, we must clarify the status of the gauge 
fields or equivalently the connections.
Given a gauge group, the transformation of gauge fields and the 
couplings with matter fields are uniquely determined.

In this paper we consider  the first order formalism 
by Zakharov-Shabat \cite{zakharov} and Ablowitz-Kaup-Newell-Segur
\cite{ablowitz}.
It uses a gauge covariant set of linear equations on a wave function. 
Since we intend to treat integrable systems in the conventional framework 
of gauge theory as far as possible we do not incorporate the spectral 
parameter unlike \cite{ablowitz} etc.
Integrable systems appear in the equation after fixing the gauge 
and using the zero curvature condition (ZCC)\cite{zakharov}. 
In this rather common approach, we add some new insights
in this paper. They are as follows.
\par
1) We have one parameter($\A$) family of nonlinear equations which only 
differ in gauge choice.
For two specific values of $\A$, they are well known integrable equations, 
{\it i.e.} the KdV($\A=1$) and the Harry-Dym (HD)($\A=-2$)  equations. 
\par
2) For any value of $\A$ the  nonlinear equations are expressed in 
terms of weak Lax equations (whose spectral parameters are zero)
and are transformed to ones invariant under the M\"{o}bius 
transformations. 
\par
3) For $\A=\pm$ integer, the transformed equations pass the Painlev\'{e} 
test and are integrable.
\par
4) Many known and new integrable systems are produced in two ways; 
first by fixing the gauge with integer values of $\A$ and 
second by making a series of gauge transformations (the Miura transformation). 
\par
In this paper we restrict our arguments in (1+1) dimensions and the
gauge group is SL(2,R) though our formulation developed in this paper 
may be applied to higher dimensions. 
Some applications of the gauge theoretical method in
higher dimensions have been developed and applied to
the W-symmetries \cite{GK}\cite{BS}.

\medskip

The paper is organized as follows.
In section 2 we give a one parameter family of nonlinear equations. These 
equations are expressed by the weak Lax equations, whose integrabilities are 
proved in Appendix B. Appendix C discusses their exact solutions. 
In section 3 we treat the KdV equation and discuss 
about an infinite number of conserved quantities in this framework.
A series of nonlinear equations is obtained by successive gauge 
transformations. (We call this series as KdV sequence). The same procedures 
are argued on the Harry-Dym (HD) equation in section 4.
Section 5 is devoted to discussions.

\section{General framework}

\indent

We start with a system of linear equations on a wave function 
covariant under some gauge group $G$ \cite{ablowitz}
\bea
D_\mu\psi&=&(\pa_\mu-A_\mu)\psi=0.\label{2.1}
\eea
Here $A_{\mu}$ is $n\times n$ matrix valued gauge field
and $\psi$ is an $n$ column vector. 
It requires an integrability condition that is 
nothing but the zero curvature condition (ZCC)\cite{zakharov}, 
\bea
F_{\mu\nu}&\equiv&\pa_{[\mu}A_{\nu]}~-~[A_\mu,A_\nu]~=~0.\label{2.2}
\eea
The system of equations \bref{2.1} and \bref{2.2} is  
invariant under the gauge transformation
\bea
A_\mu&\to&h~A_\mu ~h^{-1}+\pa_\mu h~h^{-1},~~~~~~~~~~
\psi \to h~\psi,~~~~~~~~~~h\in G.
\label{gaugetrans}
\eea
Under the infinitesimal gauge transformation of \bref{gaugetrans}
the gauge field transforms as
\bea
\D A_\mu&=&\pa_\mu\Lambda-[A_\mu,\Lambda],~~~~~~~~h~\sim~1+\Lambda.
\label{inftrans}
\eea
It is convenient to write it in a similar form as the ZCC \bref{2.2} as
\bea
F_{\D \mu}&\equiv&\D A_\mu-\pa_\mu\Lambda-[\Lambda,A_\mu]~=~0
\label{FDM}
\eea
by the following identification
\bea
\D~~&\leftrightarrow&~~\pa_\D,~~~~~~\Lambda~~~\leftrightarrow~~~A_\D.
\eea
The general solutions have pure gauge form
\bea
A_\mu&=&\pa _\mu g~g^{-1},~~~~~~~~~~\psi=g~\psi_0,~~~~~~~~~~g~\in~G
,~~~~\psi_0={\rm constant},
\label{2.3}
\eea
and are determined up to the gauge transformation, $~g\to hg~$. 
\vs

Suppose we fix some of the gauge freedom by imposing a set of 
gauge fixing conditions
\bea
\chi_j(A)&=&0.
\label{gfixing}
\eea
In a suitable choice of the gauge fixing conditions,  
the ZCC \bref{2.2} is reduced to a set of differential equations
for independent components of fields,
\bea
\phi_\A(A)=0.
\label{difeq}
\eea 
There remain residual transformations of \bref{difeq} if there exist 
transformations preserving the equation \bref{difeq}. 
The existence of such residual transformations leads to an infinite 
set of conserved currents and is a source of the integrability 
of the nonlinear differential equations.
It is known that various integrable systems are obtained by choosing the 
gauge group $G$ and the gauge fixing conditions \bref{gfixing}
depending on how the SL(2,R) subgroup is embedded in the $G$.
We discuss the above procedure in detail for a case in (1+1) space-time 
dimensions and the gauge group is SL(2,R). 

 Parametrizing the gauge fields explicitly the equations \bref{2.1} become
\be
\partial_x \psi(t,x) ~=~ A_x(t,x)\psi,~~~~~~~~~ 
A_x~=~\left(\begin{array}{cc} R & S\\ -T & -R
\end{array}\right),\label{2.7}
\ee
\be
\partial_t\psi(t,x) ~=~A_t(t,x)\psi,~~~~~~~~~ 
A_t~=~\left(\begin{array}{cc} a & b\\  c & -a
\end{array}\right) \label{2.8},
\ee 
where $\psi=(\psi_1,\psi_2)^t$ and minus signature in front of $T$ is for 
later 
convenience. 
The integrability condition (\ref{2.2}) becomes in this case
\bea
R_t+cS& =& a'-bT \nonumber\\
S_t+2bR-2Sa& =& b'\nonumber\\
T_t+2aT+2cR& =& -c' .
\label{2.9a}                     
\eea
(Here and hereafter we denote derivatives with respect to $x$ 
by primes, {\it e.g.} $b''$ means $\pa_x^2b$ etc.)  

We first fix  two of the three gauge freedoms  as
(SL(2) principal embedding)
\bea
R=0~~~~~~ and ~~~~~~~S=1, 
\label{gaugefixing}
\eea
namely
\be
\partial_x \psi^{(DS)}(t,x) ~=~\left(\begin{array}{cc} 0 & 1\\ -T & 0
\end{array}\right)\psi^{(DS)}(t,x)\equiv A_x^{(DS)}\psi^{(DS)}.
\label{3.1}
\ee
Corresponding to this gauge fixing, the first two equations of 
\bref{2.9a} 
determine $a$ and $c$ of $A_t$ 
\be
\partial_t\psi^{(DS)}(t,x) ~=~\left
(\begin{array}{cc} -{\frac{b'}{2}} & b\\ 
-Tb-{\frac{b''}{2}} & {\frac{b'}{2}}\end{array}\right)
\psi^{(DS)}(t,x)\equiv A_t^{(DS)}\psi^{(DS)}.
\label{3.2}
\ee
We call this gauge the Drinfeld-Sokolov (DS) gauge \cite{drinfeld}.  
In the DS gauge, the last equation of \bref{2.9a} is reduced to the 
following 
equation
\be
T_t=2b'T+bT'+{\frac{b'''}{2}}. \label{3.8}
\ee

We consider infinitesimal  SL(2,R) transformations consistent with 
the choice of $A_x$ in \bref{3.1}. Since we have imposed two gauge fixing 
conditions in \bref{gaugefixing}@two parameters of $\Lambda$ are fixed by 
$F_{\delta x}=0$. There remain gauge transformations which keep the 
gauge fixing conditions invariant. The form of $\Lambda$ is parametrized by 
an arbitrary function $\ep(t,x)$ as
\bea
\Lambda&=&\left(\begin{array}{cc}-{\eps'\over 2} & \eps \\ -\eps 
T-{\eps''\over 2} & {\eps' \over 2}
\end{array}\right).
\label{Lambda}
\eea
The transformations of $T$ and $b$ are determined from $F_{\D x}=
F_{\D t}=0$ as
\be
\delta T=2\eps ' T+\eps T'+{\eps'''\over 2},
\label{3.7}
\ee
\be
\delta b=\eps_t+\eps b'- b\eps'.
\label{3.7b}
\ee

The remaining gauge freedom is fixed by imposing one further
gauge fixing condition on $T$ and $b$. 
In this paper we consider a case  
\bea
T&=& \frac{b^\A}{s},
\label{GF3}
\eea
where $\A$ and $s$ are constants.
By this choice the zero curvature condition \bref{3.8} becomes
\bea
b_t&=&\frac{2+\A}{\A}~b~b'+\frac{s}{2\A}~b^{1-\A}~b'''.
\label{eqbt}
\eea
The gauge freedoms of SL(2,R) are three and they are all fixed by choosing a 
specific value of $\A$.  
For each $\A$ there still exist transformations \bref{3.7b} 
under which \bref{eqbt} remains invariant
when there exist local solutions $\ep$ satisfying
\bea
\ep_t&=&
\frac{2+\A}{\A}~b~\ep'+\frac{s}{2\A}~b^{1-\A}~\ep'''.
\label{eqept}
\eea
Every local solutions of \bref{eqept}
correspond to the residual transformations for \bref{eqbt}. 

We will examine the integrability
of \bref{eqbt}. 
First we can select integrable systems by the Painlev\'{e} test \cite{weiss} 
though it is a sufficient condition but not a necessary one.
One can check that only the case $\A =1$  passes the Painlev\'{e} test 
(the value of $s$ is irrelevant and we take $s=2$ for convenience) .
In this case $b=2T$ and  \bref{eqbt} gives the KdV equation, which will be 
discussed in detail in the next section. 

Next we consider \bref{eqbt} in the weak Lax equation \cite{nucci}.
As is easily checked, \bref{eqbt} has the weak Lax pairs, $L_1$ and $L_2$,
\be
\psi''+\frac{b^{\A}}{s}\psi\equiv  L_1\psi=0,
\label{eqL}
\ee
\be
\psi_t-b\psi'+\frac{1}{2}b'\psi\equiv L_2\psi=0.
\label{eqT}
\ee
Indeed, $[L_1,L_2]\psi =0$ gives \bref{eqbt} for arbitrary $\A$.

\par
Let us define $\varphi$ by $\varphi\equiv\frac{\psi_1}{\psi_2}$, 
where $\psi_i$ are two linearly independent solutions of \bref{eqL} and \bref{eqT}. The concrete example of $\varphi$ is discussed in Appendix A.  
Then $\varphi$ satisfies
\be
(\frac{\varphi_t}{\varphi '})^\A=\frac{s}{2}\{\varphi;x\},
\label{eqA}
\ee
where $\{\varphi;x\}$ is the Schwarzian derivative,
\be
\{\varphi;x\}\equiv \frac{\varphi '''}{\varphi '}-
\frac{3}{2}\frac{\varphi ''^2}{\varphi '^2}.
\ee
\bref{eqA} is invariant under the M\"{o}bius transformation by construction,
\be
\varphi\rightarrow \frac{a+b\varphi}{c+d\varphi},\hspace{1 cm}ad-bc=1.
\ee
Using the transformation of dependent variable,
\be
\varphi =e^ F,
\ee
the equation \bref{eqA} is transformed to
\be
\left(\frac {F_t}{F'}\right)^{\A}-\frac{s}{2}\left(\{F;x\}-\frac{F'}{2}^2
\right)=0.
\label{eqF}
\ee
Here we introduce new dependent variables \cite{lou}
\be
F'\equiv u, \hspace{2 cm}F_t\equiv v.
\label{lou}
\ee
Substituting \bref{lou} into \bref{eqF}, we obtain a coupled equations
 of $u$ and $v$,
\be
\left(\frac{v}{u}\right)^{\A}-\frac{s}{2}\left(\frac{u''}{u}-
\frac{3}{2}\left(\frac{u'}{u}\right)^2-\frac{u^2}{2}\right) =0
\label{lou1}
\ee
and
\be
u_t=v'.
\label{lou2}
\ee
We apply the Painlev\'{e} test to the coupled equations, \bref{lou1} and 
\bref{lou2}. We find, for $\A=\pm $ integer, that they universally 
have the resonances -1,1,1 and pass the Painlev\'{e} test. 
The Painlev\'{e} test and the exact solutions of these equations are discussed in Appendices B and C.  
Thus it follows \bref{eqbt} is integrable for any integer $\A$.
For $\A=-2, s=1$, \bref{eqbt} gives 
the HD equation and \bref{eqept} gives its residual transformations. 
The HD equation is a well known example which does not pass the 
Painlev\'{e} test in its original form.  
However, as we mentioned, the HD equation expressed in a coupled 
system \bref{lou1} and \bref{lou2} passes the Painlev\'{e} test
and is integrable.
Some related matters with the HD equation will be discussed in section 4. 
The equations for the other integer $\A$ are new integrable systems.

\section{Gauge Transformation and the KdV Sequences}
\indent

The equation corresponds to $\A =1$ (and $s=2$) in \bref{GF3} and \bref{eqbt} gives the KdV equation,
\be
T_t=T'''+6TT'\label{3.9}.
\ee
\bref{eqA} becomes
\be
\frac{f_t}{f'}=\{f;x\},
\label{SKdV}
\ee
where $f$ is $\varphi$ for the KdV (see Appendix A) and $T$ is related with $f$ by
\be
T=\frac{1}{2}\{f;x\}.
\label{T}
\ee
\bref{SKdV} is called Schwarz KdV (SKdV) equation \cite{wilson},
\cite{hereman}. 
\bref{eqept} gives the residual transformations for the KdV equation, 
\bea
\eps_t&=&\eps '''~+~6~T~\eps '.
\label{eqep}
\eea
This is {\it the linear equation associated to the KdV 
equation} which Gardner et.al obtained from the inverse scattering 
method\cite{gardner}. 

The equation \bref{eqep} has an infinite number of solutions for 
the parameter function $\ep$, which lead to an infinite number of conserved 
quantities of the KdV system.
It is shown as follows. In writing 
\bea
\eps ' &=& \tilde\eps
\eea
$\tilde\eps$ satisfies, from \bref{eqep} 
\bea
\tilde\eps_t&=&\tilde\eps'''~+~6~T~\tilde\eps '~+~6~T'~\tilde\eps.
\label{eqtep}
\eea
We can show $\D T$ of \bref{3.7} is a solution of \bref{eqtep}
\bea
\tilde\eps&=&\D T(\eps)
\label{tildeep}
\eea
when $T$ and $\eps$ satisfy \bref{3.9} and \bref{eqep} respectively.
We start from a solution of  \bref{eqep}, 
$\eps^{(0)}=1$.
We find a solution of \bref{eqtep} as $\tilde\eps^{(1)}=\D T(\eps^{(0)})
=
T'$. By integrating it 
we find second solution of  \bref{eqep}, $\eps^{(1)}=T$. Repeating
this procedure an infinite set of residual transformations $\eps^{(n)}$ of the 
KdV equation is obtained in local forms. They are related with an infinite 
number of conserved densities \cite{wadachi}.  
Wadachi showed a theorem:
If $\phi_x$ satisfies
\be
\phi_{xt}=K(\phi_x)
\ee
and if it has an infinite conserved density ${\cal G}^{(n)}$, then $\frac{\D 
{\cal G}^{(j)}}{\D \phi_x}$ satisfy
\be
\int_{-\infty}^{\infty}dx [h(x)\frac{\partial}{\partial t}\frac{\D {
\cal G}^{(j)}}{\D \phi_x}+\frac{d}{d\eps}K(\phi_x+\eps h)\frac{\D 
{\cal G}^{(j)}}{\D \phi_x}]=0.
\label{wadachi}
\ee
For the KdV equation,
\be
K(\phi_x)=6\phi_x\phi_{xx}+\phi_{4x}
\ee
and $\frac{\D {\cal G}^{(j)}}{\D \phi_x}$ satisfies the same equation 
\bref{eqep} for $\eps$. So $\eps^{(j)}$ are related with $j$-th conserved
density ${\cal G}^{(j)}$ by
\be
\frac{\D {\cal G}^{(j)}}{\D \phi_x}=\eps^{(j)}.  
\ee
\bref{tildeep} is another interpretation of the bi-Hamiltonian relation 
\cite{blaszak},
\be
\pa_x \frac{\D {\cal G}^{(j)}}{\D \phi_x}=(\pa_x T+T\pa_x +\frac{\pa_x^3}{2}) 
\frac{\D {\cal G}^{(j-1)}}{\D \phi_x}.
\label{bihamilton}
\ee 

Next we consider the Miura transformations from the gauge 
theoretical view point.
Retaining $b=2T$, we take the following gauge transformation from DS 
gauge.
Since we have imposed three gauge fixing conditions 
there remains no gauge degree of
freedom. Any gauge transformations \bref{gaugetrans} change the forms of
gauge fixing conditions.
Suppose we make a finite transformation  \bref{gaugetrans} by ($h=V^{(M)}$)
\bea
\psi^{(M)}(t,x)=~\left(\begin{array}{cc} 1, & 0\\ -j(t,x), & 1
\end{array}\right) \psi^{(DS)}(t,x)\equiv V^{(M)}\psi^{(DS)}.
\label{4.1}
\eea
Under this transformation the gauge fields are mapped 
from those in the DS gauge to
\bea
A_x^{(M)} ~&=&~
(\partial_x V^{(M)}(t,x)~+~V^{(M)}(t,x)A_x)V^{(M)}(t,x)^{-1}\nonumber\\ 
     & = & ~\left(\begin{array}{cc} j, & 1\\ -T-j^2-j', & -j
\end{array}\right),\label{4.4}
\eea
\bea
A_t^{(M)} ~&=&~
(\partial_t V^{(M)}(t,x)~+~V^{(M)}(t,x)A_t)V^{(M)}(t,x)^{-1}\nonumber\\ 
     & = & ~\left(\begin{array}{cc} -T'+2Tj, & 2T\\ 
-j_t-2T^2-2Tj^2+2jT'-T'', & T'-2Tj
\end{array}\right).\label{4.5}
\eea
The transformation function $\Lambda$ in \bref{Lambda}, which kept the
gauge fixing conditions \bref{gaugefixing} invariant, is also mapped to
\bea
\Lambda^{(M)}&=&(\D V^{(M)}~+~V^{(M)} 
\Lambda)V^{(M)}(t,x)^{-1}\nonumber\\ 
     & = & ~\left(\begin{array}{cc} -{\ep'\over 2}+\ep j, & \ep\\ 
-\D j-\ep T-\ep j^2+j\ep'-\frac{\ep''}{2}, & {\ep'\over 
2}-\ep j
\end{array}\right).\label{4.5b}
\eea
They are solutions of $F_{xt}=F_{\D x}=F_{\D t}=0$ in the new gauge.

So far $j(t,x)$ in $V^{(M)}$ is not specified. We choose it as
\be
T=-j^2-j'\label{4.2}
\ee
so that the $(A_x^{(M)})_{21},~$ (2,1) component of matrix $A_x^{(M)}$,~ 
vanishes.
The ZCC condition requires $(A_t^{(M)})_{21}$ to be zero also and
\bea
j_t&=&-6j^2j'+j'''.                        
\label{MKdV}
\eea
This is the well-known MKdV equation.  
The compatibility requires that $(\Lambda^{(M)})_{21}$ in the gauge 
transformation \bref{4.5b}
must vanish and
\bea
\D j~=~j\ep'+j'\ep-\frac{\ep''}{2}.                          
\label{delj}
\eea
This is the transformation property of $T$ in \bref{3.7} in terms of 
$j$.
From \bref{T} and \bref{4.2} 
\be
j=-\frac{f''}{f'}.
\label{fj}
\ee

We can repeat the  same form of gauge transformation as \bref{4.1} on
the MKdV equation furthermore and get other integrable systems;
\bea
\psi^{(C)}= ~\left(\begin{array}{cc} 1 & 0\\ -\eta (t,x) & 1
\end{array}\right) \psi^{(M)}(t,x)\equiv V^{(C)}\psi^{(M)},
\eea
where $\eta$ is chosen as $(A_x^{(C)})_{21}$ in the new gauge vanishes 
\be
\eta'=-\eta^2-2j\eta.
\ee
The ZCC determines the differential equation for $\eta $ with $b=2T$,
\be
\eta_t=\eta '''-{1\over 2}(3\eta'^2\eta^{-1}+\eta^3)_x.
\ee
This is the Calogero Korteweg-de Vries (CKdV) equation \cite{calogero}.
The successive B\"{a}cklund transformations from the KdV to CKdV equations 
were discussed in the bilinear formalism in \cite{nakamura}.
The residual symmetry transformation of $\eta $ is 
\bea
\D\eta &=& (\eps\eta)_x.
\eea
$\eta$ is also related with $f(t,x)$ of the SKdV equation by
\be
\eta=\frac{f'}{f}.
\label{feta}
\ee
Thus by the successive gauge transformations of the type 
\bref{4.1}, we get a series of integrable equations, 
which we refer as the KdV sequence.  
The same series of integrable equations have been obtained 
by expanding the (usual) Lax equation in power series of the 
spectral parameter \cite{pavlov}.  
Namely we begin with
\be
\psi''+T\psi=\lambda \psi
\ee
and expand $\psi$ as
\be
\psi=exp(\int dx(r^{(0)}+r^{(1)}\lambda +r^{(2)}\lambda^2+....)).
\ee
Then the recursion equation of $r^{(k)}$ gives the same series of 
integrable equations.  
Thus the successive Miura transformations with no spectral parameter in our theory correspond to the expansion of the wave function by the spectral parameters in the conventional theories.

\section{Harry Dym Equation}
\indent

The Harry Dym (HD) equation \cite{kruskal},
\be
b_t=-{1\over 4}b^3b''',
\label{HD}
\ee
corresponds to $\A =-2$ (and $s=1$) in \bref{eqbt}.
In this gauge \bref{3.1} and \bref{3.2} become 
\be
\partial_x \psi^{(HD)}(t,x) ~=~\left(\begin{array}{cc} 0 & 1\\ -{1\over 
b^2} & 0
\end{array}\right)\psi^{(HD)}(t,x)\equiv A_x^{(HD)}\psi^{(HD)},
\label{HDgaugex}
\ee
\be
\partial_t\psi^{(HD)}(t,x) ~=~\left(\begin{array}{cc} -{b'\over 2} & b\\ 
-{1\over b}-{b''\over 2} & {b'\over 2}
\end{array}\right)\psi^{(HD)}(t,x)\equiv A_t^{(HD)}\psi^{(HD)}.
\label{HDgauget}
\ee

The residual transformation $\Lambda$ in this case is
\be
\Lambda^{(HD)}=~\left(\begin{array}{cc}-{\eps'\over 2} & \eps \\ 
-{\eps\over 
b^2}-{\eps''\over 2} & {\eps' \over 2}
\end{array}\right)
\ee 
and
\be
\delta b=-b\eps '+b'\eps -{b^3\over 4}\eps '''.
\label{deltaB}
\ee
\bref{HD} has the residual gauge transformations, that is, \bref{deltaB} with 
$\eps$ restricted by
\be
\eps_t=-{1\over 4}b^3\eps '''.
\label{ept}
\ee
Applying \bref{wadachi} in the HD equation, we obtain the linear equation 
associated to the HD equation
\bea
\veps_t~=~\frac34 b^2b'''\veps~-~\frac14(b^3\veps)'''
\label{Aslinear},
\eea
where
\bea
\veps^{(j)}&=&\frac{\D {\cal G}^{(j)}}{\D b}
\eea
with ${\cal G}^{(j)}$ j-th conserved density. It
 should be remarked
 that in the HD case the associated
 linear equation \bref{Aslinear} has different form from the equation of 
residual gauge transformation \bref{ept} unlike the case of KdV.
We can check \bref{Aslinear}
for first several terms:
\bea
{\cal G}^{(0)}&=&\frac{-1}{b},~~~~~~~~~~
{\cal G}^{(1)}~=~\frac{b'^2}{2b},~~~~~~~~~~
{\cal G}^{(2)}~=~\frac{b'^4}{8b}+\frac{bb''^2}{2},\nonumber\\
{\cal G}^{(3)}&=&\frac{b'^6}{16b}+\frac{3bb'^2b''^2}{4}-
\frac{b^2b''^3}{2}+\frac{b^3b'''^2}{2},... .
\eea
 $\veps$ in \bref{Aslinear} and $\ep$ in \bref{ept} are related by
\bea
\ep~=~b^3~\veps.
\label{relee}
\eea

$\veps$'s satisfy the bi-Hamiltonian relations like \bref{bihamilton} in the KdV case,
\bea
-b^2~\pa_x~b^2~\veps^{(n)}&=&b^3~\pa_x^3~b^3~\veps^{(n-1)}
\eea
from which the conserved densities ${\cal G}^{(n)}$ follow.
\bref{relee} tells that series of solutions of \bref{ept} is obtained from
\bea
-~\pa_x~(\frac{\ep^{(n)}}{b})&=&b~\pa_x^3~\ep^{(n-1)}.
\eea
Namely starting from $n=0$ and $\eps^{(-1)}=1$ we obtain 
\bea
\ep^{(-1)}=1,~~~~~~~
\ep^{(0)}={b},~~~~~~~
\ep^{(1)}=b(\frac{b'^{2}}2-bb''),~~~~~~~
\nonumber\\
\ep^{(2)}={\frac{3\,b\,{{b'}^4}}{8}} - 
  {\frac{3\,{{b }^2}\,{{b'} }^2\,
      b''} {2}} + 
  {\frac{3\,{{b }^3}\,{{b''} }^2}{2}} + 
  2\,{{b }^3}\,b' \,b'''  +   {{b }^4}\,b^{(4)},... ~. 
\eea
In this way we can 
obtain local expressions for $\ep^{(n)}$ and 
give an infinite residual symmetry of HD equation through
\bref{deltaB} and \bref{ept}. 
\par

Repeating the same procedure as in the previous section, we can get the 
HD sequence as follows.
Retaining $T={1\over b^2}$, 
we perform the gauge transformations from the DS gauge and obtain, 
so called, the HD sequence as follows.
\bea
\psi^{(mHD)}(t,x)=~\left(\begin{array}{cc} 1 & 0\\ -k(t,x) & 1
\end{array}\right) \psi\equiv V^{(mHD)}\psi.
\eea

The gauge fields are transformed from \bref{HDgaugex} and \bref{HDgauget}
 to
\bea
A_x^{(mHD)} ~&=&~
(\partial_x 
V^{(mHD)}(t,x)~+~V^{(mHD)}(t,x)A_x^{HD})V^{(mHD)}(t,x)^{-1}\nonumber\\ 
     & = & ~\left(\begin{array}{cc} k, & 1\\ -k'-k^2-{1\over b^2}, & -k
\end{array}\right)
\eea
and
\bea
A_t^{(mHD)} ~&=&~
(\partial_t 
V^{(mHD)}(t,x)~+~V^{(mHD)}(t,x)A_t^{HD})V^{(mHD)}(t,x)^{-1}\nonumber\\ 
     & = & ~\left(\begin{array}{cc} -{b'\over 2}+bk, & b\\ 
-k_t+b'k-{1\over b}-{b''\over 2}-bk^2, & -bk+{b'\over 2}
\end{array}\right).
\eea
Here we apply the same rule as in the KdV sequence also that the (2,
1) components of gauge transformed $A_{\mu}$ are zero:
\be
{1\over b^2}=-k'-k^2,
\label{mHD1}
\ee
\be
-k_t+b'k-{1\over b}-{b''\over 2}-bk^2=0
\label{mHD2}
\ee
and
\bea
\D k&=&\eps'k+\eps k'-{\eps''\over 2}.
\eea
In this way we can obtain a series of non-linear equations analogous to the 
KdV sequences, though we do not repeat the arguments furthermore.
The explicit proof of integrability of this series apart from the HD 
equation is still open though it is formulated in the same ways as the 
KdV series in our formalism. 
Antonowicz and Fordy discussed different extensions of the HD equation 
in the inverse scattering framework \cite{antonowicz}

\section{Discussion}
\indent

In this paper we have analyzed several integrable systems in the 
framework of gauge theory. Starting with the DS gauge, we have fixed the 
remaining gauge freedom by (2.20) and obtained the one parameter ($\A$) 
family of nonlinear equations. 
It was shown explicitly for $\A=1$ (KdV) and $\A=-2$ (HD) cases that
the residual invariant transformations  assure 
an infinite number of conserved quantities.  The relations between the 
residual 
transformations and conserved quantities are simple for the KdV but are
not for the others. The one parameter family of equations (2.21) was  
discussed in the framework of weak Lax equation. The set of 
equations \bref{lou1} and \bref{lou2} was shown to pass the Painlev\'{e} 
test for $\A =\pm$ integer and are integrable.   
Applying the gauge transformations (the Miura transformations) successively 
to (2.14) and (2.15) with (2.20), we have also constructed the KdV and HD 
sequences. 

It is very interesting to discuss other typical integrable systems like 
the Burgers, Sawada-Kotera, Kaup-Kuperschmidt, and Sine-Gordon equations etc. 
in this formalism. Some of them, at least, can be incorporated in our 
formalism by relaxing the DS gauge. However, important is not the fact 
that our formalism simply invokes these known and unknown integrable systems 
but is that those equations are mutually related by the gauge fixings 
($\A =\pm$ integer) and gauge transformations (the Miura transformations 
for each $\A$) in the gauge theoretical framework. 
It is our goal to understand integrable systems in the framework of gauge 
theories in a self complete manners.  
In this paper we have partially succeeded in it but there still  left
many points unanswered. For instance, we have not obtained an infinite number 
of conserved quantities for all integer $\A$. 
Although in this paper we have restricted our arguments in (1+1) dimensions 
and SL(2,R) gauge group, they can be applied to higher 
dimensions and to other gauge groups, for example W-symmetries.
\vspace{2 cm}

\appendix

\section{Example of $\varphi\equiv\psi _1/\psi _2$}

Let us consider the two linerly independent solutions of Eq.(2.23) for the KdV equation.  
Eq.(2.14) reads
\be
\psi ''(x,t) +T(x,t)\psi(x,t)=0.
\label{gs1}
\ee
If we put
\be
\psi~=~e^{g(x,t)},\label{gs2}
\ee
it is a solution of (A1) if T(x,t) is given by
\be
T(x,t)~=~-g'(x,t)^2-g''(x,t).\label{gs3}
\ee
For another solution, we must solve
\be
\psi(x,t)''~+~ (-g'(x,t)^2-g''(x,t))\psi(x,t)~=~0 \label{gs4}
\ee
for given $g(x,t)$~. It has a trivial solution $~\psi=A~e^{g(x,t)}~$ 
for $x$ independent $A$. We look for solution with $x$ dependent $A$,
$$~\psi~=~A(x,t)~e^{g(x,t)}.~$$
The equation of motion for $A$ is obtained from Eq.(A.1). 
\be
A''~+~2 ~g'(x,t) A'~=~0, \label{gs5}
\ee
which can be integrated as
\be
A'(x,t)~=~c_1(t)~e^{-2g(x,t)},~~~\longrightarrow~~~
   A(x,t)~=~c_2(t)+~c_1(t)~\int^x dx'~e^{-2g(x',t)}.\label{gs6}
\ee
Thus the solution is 
\be
\psi(x,t)~=~(c_2(t)~+~c_1(t)~\int^x dx'~e^{-2g(x',t)})~e^{g(x,t)},~~~~~
T(x,t)~=~-g'(x,t)^2-~ g''(x,t).\label{gs7}
\ee
It is the general solution containing one arbitrary function of  
$g(x)$ and two constants $c_1,~c_2$ with respect to $x$.
\vskip 10mm

The solution contains an integral expression. In place of $g(x,t)$ we 
can use 
\be 
f(x,t)~=~\int^x dx'~e^{-2g(x',t)} \label{gs8}
\ee
as the arbitrary function describing the solution. In terms of $f(x,t)$, 
the general solution has the local form;
\be
\psi~=~f'(x,t)^{-1/2}~(c_1~f(x,t)~+~c_2),~~
T(x,t)~=~\frac12~(~{f'''(x,t)\over  f'(x,t)}-\frac32{f''^2(x,t)\over 
f'(x,t)^2})\equiv {1\over 2}\{f(x,t);x\}.
\ee
Thus the two linearly independent solutions, for instance, are 
\be
f'(x,t)^{-1/2}~(c_1~f(x,t)~+~c_2)\equiv c_1\psi_1(x,t)+c_2\psi_2(x,t)
\ee
and 
\be
\varphi\equiv \frac{\psi_1}{\psi_2}=f(x,t)
\ee
The two linearly independent solutions in general are given by
\be
\Psi_1=a\psi_1+b\psi_2,~~\Psi_2=c\psi_1+d\psi_2
\ee
with $ad-bc=1$ (please do not confuse the constants $a,\sim,d$ with the functions $a,\sim,d$ in Eq.(2.11)). In this case $\varphi$ changes as Eq.(2.27) but Eq.(3.2) is invariant as it should be.

\section{ Painlev\'{e} test of \bref{lou1} and \bref{lou2}}
We apply the Painlev\'{e} test to the coupled equations, \bref{lou1} and 
\bref{lou2} for $\A=$ an arbitrary integer. 
As will been shown, the resonances are -1,1 and 1 for any integer
$\A$.  First we consider the case of $\A=$ positive 
integer in which the KdV equation ($\A=1$) is included, and second the case of 
$\A=$ negative integer to which the HD equation ($\A=-2$) belongs.
\subsection{Case of $\A=$ positive integer}
For the case of $\A=$positive integer ($\equiv n$), \bref{lou1} is 
transformed to
\be
v^n-\frac{s}{2}\left(u''u^{n-1}-
\frac{3}{2}u'^2u^{n-2}-\frac{u^{n+2}}{2}\right)=0.
\label{P1}
\ee
As usual we expand $u$ and $v$ around a
movable pole $\gamma^{-1}$. To obtain the leading order term, substitute
\be
u=u_0\gamma^i ~~~~~~v=v_0\gamma^j
\ee
with $u_0v_0\neq 0$ into \bref{P1} and \bref{lou2}.
Then \bref{lou2} gives $i=j$ and the dominant equation of \bref{P1} with
 this result gives i=j=-1, and
\be
u_0=\pm\gamma_x~~~v_0=\pm\gamma_t ~~\rm{(double~ sign~ is~ in ~same~ order)}
\label{leading}
\ee
irrelevantly to $n$.
(It should be remarked that the dominant terms of \bref{P1} are the terms 
in the bracket.)  So we can expand $u$ and $v$ as
\be
u=\sum_{i=0}^{\infty}u_i\gamma^{i-1},~~~v=\sum_{j=0}^{\infty}v_j\gamma^{j-1}.
\label{expansion}
\ee
Substitute \bref{expansion} into  \bref{P1} and \bref{lou2}, 
perturb up 
to O($\gamma^{j-n-2}$) and O($\gamma^{j-2}$), respectively, and pick up 
$u$, $v$ having the largest suffix j, then we get the resonance equation:
\be
\left(
\begin{array}{cc}
(j+1)(j-1)u_0\gamma_x^2& 0 \\
(j-1)\gamma_t&-(j-1)\gamma_x\\
\end{array}
\right)
\left(\begin{array}{c}u_j\\v_j
\end{array}
\right)=\left(\begin{array}{c}f(u_i,v_i;i\leq j)\\u_{j-1,t}-v_{j-1,x}
\end{array}
\right).
\ee
Thus we have the resonances, -1,1,1.

We can check the consistency
by
substituting \bref{expansion} into  \bref{P1} and 
collecting the terms of order $\gamma^{-(n+1)}$, we find 
\be
2(n-1)u_1u_0^{n-1}\gamma_x^2-\frac{3}{2}(n-2)u_1u_0^{n-1}\gamma_x^2-
\frac{1}{2}(n+2)u_1u_0^{n+1}\equiv 0
\ee
with the help of \bref{leading}.  
Thus $u_1$ and, therefore, $v_1$ from \bref{lou2} are indefinite.  
The subsequent coefficients $u_i$,~$v_i ~(i\geq 2$) are definite by 
virtue of the first term of \bref{P1}.
Thus \bref{lou1} and \bref{lou2} have passed
 the Painlev\'{e} test.
\subsection{Case of $\A=$ negative integer}
For the case of $\A=$ negative integer ($\equiv -n$), \bref{lou1} is transformed to
\be
u^n-\frac{s}{2}\left(\frac{u''}{u}-
\frac{3}{2}\left(\frac{u'}{u}\right)^2-\frac{u^2}{2}\right)v^n =0
\label{P3}
\ee
In this case the Painlev\'{e} test is performed analogously to the positive integer case or even simpler than before since the resonance equation is determined by the factor in front of $v^n$ in \bref{P3}.
So we do not repeat the argument.

\vspace{1 cm}

\section{Exact solutions to the system of \bref{lou1} and \bref{lou2}}
To obtain the exact solutions to the system of \bref{lou1} and \bref{lou2} we change the dependent variables as
\be
u=a({\rm ln}~f)_x~~~v=a({\rm ln}~f)_t,
\label{change}
\ee
where we have took into consideration \bref{lou2} and the fact that $u$ and $v$ have the leading order -1.
We search the exact solutions for each specific value of $\A$.
First we consider the case $\A=1$.
Substituting \bref{change} into \bref{lou1} and setting $a=1$ 
to eliminate the terms of order $f^{-4}$, we obtain 
\be
4f_tf'+3sf''^2-2sf'f'''=0
\label{bi}
\ee
This equation is same as \bref{SKdV} as is expected.
This equation can not be written as the bilinear form \cite{hirota}
but we can obtain the dispersion relation and one soliton solution \'{a} la Hirota Method. Namely let us set $f(t,x)=1+{\rm{exp}}(px+qt)$ in \bref{bi}, then the dispersion relation is 
\be
q=-\frac{s}{4}p^3.
\label{dispersion1}
\ee
Therefore it goes from \bref{change} and \bref{bi} that
\bea
u&=&\frac{p}{2}e^{\zeta/2}\rm{sech}~(\zeta/2)\\
v&=& \frac{q}{2}e^{\zeta/2}\rm{sech}~(\zeta/2)
\eea
with
\be
\zeta\equiv px+qt.
\ee
Formally we can trace back this one soliton solution to those of the KdV sequence by the relations with $f$ like \bref{T}, \bref{fj}, \bref{feta} etc.
Unfortunately, this process gives trivial (constant) solutions  for the KdV and MKdV equations. However, \bref{feta} gives the solution for the CKdV equation as
\be
\eta =\frac{p}{2}e^{\zeta/2}{\rm{sech}}~(\zeta/2) =u.
\ee 
\bref{lou1} and \bref{lou2} does not allow multi-soliton solutions. Namely let us put
\be
f_2=1+\epsilon (e^{p_1x+q_1t}+e^{p_2x+q_2t})+\epsilon^2A_{12}e^{(p_1+p_2)x+(q_1+q_2)t}
\ee
for two solitons solution and substitute it into \bref{SKdV}. Equating coefficients of like powers of $\epsilon$ to zeros, we obtain 
\be
q_i=-\frac{s}{4}p_i^3  ~~(i=1,2)
\ee
for O(1),
\be
-3s{\rm exp}\{-\frac{s}{4}(p_1^3+p_2^3)t+(p_1+p_2)x\}p_1(p_1-p_2)^2p_2=0
\ee
for O($\epsilon$), leading to $p_1=p_2$, and
\be
12sA_{12}^2{\rm exp}(-sp_1^3t+4p_1^4x)p_1^4=0
\ee
for O($\epsilon^4$). O($\epsilon^2$) and O($\epsilon^3$) terms vanish 
identically.  Obviously these equations are not satisfied for $p_i\neq 0$.
\par
For the other values of $\A$, as is easily seen from the previous arguments, the situation is not changed except for replacing the dispersion \bref{dispersion1} by
\be
q=(\pm )^{\A +1}p(-\frac{s}{4}p^2)^{1/\A}.
\label{dispersion2}
\ee
There is no multi solitons solution.  This is a natural consequence from 
the fact that \bref{lou1} and \bref{lou2} can not be written in bilinear forms.
 However, as was shown in Appendix B this system of equations is integrable.

\vspace{1 cm}
\par


\begin{thebibliography} {99}
\bibitem{das}
For the review of integrable systems see, for instance, A Das,  
Integrable Models (World Scientific, 1989 );
V.E. Zakharov (ed.), What is integrability ? (Springer-Verlag, 1990)
\bibitem{ward}
R.S. Ward, Philos. Trans. R. Soc. London, Ser. {\bf A 315} 451-457 (1985)
\bibitem{zakharov}
V. E. Zakharov and A. B. Shabat, JETP {\bf 34} 62-69 (1972)
\bibitem{ablowitz}
M. J. Ablowitz, D. J. Kaup, A. C. Newell and H. Segur,  Phys. Rev. Lett. 
{\bf 30} 1262-1264 (1973); ibid. {\bf 31} 125-127 (1973)
\bibitem{GK}
J.Gomis, J.Herrero, K.Kamimura and J.Roca, 
Prog. Theor. Phys. {\bf 91}413-418 (1994), 
 Ann. of Phys. {\bf 244} 67-100 (1995),
 Phys. Lett. {\bf B339} 59-64 (1994)
\bibitem{BS}
For reviews of W-symmetry, see, for instance, L.Feh\'{e}r, L. O'Raifeartaigh, 
P. Ruelle, I. Tsutsui and A. Wipf, Phys. Rep. {\bf 222} 1-64 (1992);
P. Bouwknegt and K. Shoutens, Phys. Rep. {\bf 223} 183-276 (1993)
\bibitem{drinfeld}
V. Drinfeld and V. Sokolov, J. Sov. Math. {\bf 30} 1975-2036 (1984)
\bibitem{weiss}
J. Weiss, U. Tabor and G. Carneval,  J. Math. Phys. {\bf 24} 522-526 (1983)
For a Review of the Painlev'{e} test see, for instance,  A. Ramani, 
B. Grammaticos and T. Bountis ,  Phys. Rep. {\bf 180} 159-245 (1989)
\bibitem{nucci}
M.C. Nucci, J. Phys. A {\bf 22} 2897-2913 (1989)
\bibitem{lou}
S. Lou, J. Math. Phys. {\bf 39} 2112-2121 (1988)
\bibitem{wilson}
G. Wilson, Phys. Lett. {\bf A132} 445-450 (1988)
\bibitem{hereman}
W. Hereman, P.P. Banerjee and M.R. Chatterjee, J. Phys. {\bf A 22} 
241-255 (1989)
\bibitem{gardner}
C.S. Gardner, J.M. Green, M.D. Kruskal and R.M. Miura, Comm.Pure Appl. M
ath. {\bf 27} 97-133 (1974)
\bibitem{wadachi}
M. Wadachi, Stud. Appl. Math. {\bf 59} 153-186 (1978)
\bibitem{blaszak}
See,for instance, M. Blaszak, Multi-Hamiltonian Theory of Dynamical 
Systems (Springer, 1998) 
\bibitem{calogero}
F. Calogero and A. Degasperis, J. Math. Phys. {\bf 22} 23-31 (1981) 
\bibitem{nakamura}
A. Nakamura and R. Hirota, J. Phys. Soc. Jpn. {\bf 48} 1755-1762 (1980)
\bibitem{pavlov}
M.V. Pavlov, Phys. Lett. {\bf A243} 295-300 (1998)
\bibitem{kruskal}
M.D. Kruskal, Lecture notes in physics, {\bf 38} (Springer-Verlag, 
Berlin, 1975)
\bibitem{antonowicz}
M. Antonowicz and A.P. Fordy, J. Phys. {\bf A 21} L269-L275 (1988)
\bibitem{hirota}
R. Hirota,  Mathematical Theory of Solitons in Japanese (Iwanami, Tokyo, 1992)
\end{thebibliography}
\end{document}